\documentclass[12pt,aps,prl,amsmath,amssymb,sort&compress,comma,super,nofootinbib,floatfix]{revtex4-1}
\usepackage{graphicx}
\usepackage{bm}
\begin{document}

\title{Decrease of d-wave pairing strength in spite of the persistence of magnetic excitations in the overdoped Hubbard model}

\author{Edwin W. Huang}
\affiliation{Department of Physics, Stanford University, Stanford, California 94305, USA}
\affiliation{Stanford Institute for Materials and Energy Sciences, SLAC National Accelerator Laboratory and Stanford University, Menlo Park, CA 94025, USA}
\author{Douglas J. Scalapino}
\affiliation{Department of Physics, University of California, Santa Barbara, CA 93106-9530, USA}
\author{Thomas A. Maier}
\affiliation{Computational Sciences and Engineering Division and Center for Nanophase Materials Sciences,
Oak Ridge National Laboratory, Oak Ridge, Tennessee 37831-6494, USA}
\author{Brian Moritz}
\affiliation{Stanford Institute for Materials and Energy Sciences, SLAC National Accelerator Laboratory and Stanford University, Menlo Park, CA 94025, USA}
\author{Thomas P. Devereaux}
\affiliation{Stanford Institute for Materials and Energy Sciences, SLAC National Accelerator Laboratory and Stanford University, Menlo Park, CA 94025, USA}

\date{\today}

\begin{abstract}
Evidence for the presence of high energy magnetic excitations in overdoped La$_{2-x}$Sr$_x$CuO$_4$ (LSCO)
has raised questions regarding the role of spin-fluctuations in the pairing mechanism. If they remain present in overdoped LSCO, why does $T_c$ decrease in this
doping regime? Here, using results for the dynamic spin susceptibility ${\rm Im}\chi(\bm{q},\omega)$
obtained from a determinantal quantum Monte Carlo (DQMC) calculation for the Hubbard model
we address this question. We find that while high energy magnetic excitations persist in the overdoped
regime, they lack the momentum to scatter pairs between the anti-nodal regions. It is the decrease in
the spectral weight at large momentum transfer, not observed by resonant inelastic X-ray scattering (RIXS), which
leads to a reduction in the $d$-wave spin-fluctuation pairing strength.
\end{abstract}


\maketitle

Recent resonant inelastic X-ray scattering (RIXS) studies of La$_{2-x}$Sr$_x$CuO$_4$
(LSCO) have found that high energy magnetic excitations near the antiferromagnetic zone boundary are present across a wide range of doping in the LSCO phase diagram \cite{Dean2013,Wakimoto2015,Meyers2017}. In particular, while these excitations gradually soften and broaden in the overdoped region, they remain even as the
superconducting transition temperature $T_c$ decreases. This raises questions
regarding the role of spin fluctuations in the pairing mechanism \cite{Scalapino2012}. Specifically,
if these magnetic excitations persist in the overdoped LSCO, what is
responsible for the destruction of high temperature superconductivity?

Here we discuss results for the dynamic spin susceptibility ${\rm Im}\chi(\bm{q},\omega)$,
obtained from determinantal quantum Monte Carlo (DQMC) calculations for the doped
2D Hubbard model \cite{White1989,Jia2014,Kung2015}. We find that similar to the RIXS studies, high-energy magnetic excitations persist into the overdoped regime. However at large momentum transfer, beyond the range observed
by RIXS \cite{Ament2011}, a reduction and hardening of the strength of the spin-fluctuation spectral weight is observed. We discuss the doping dependence of magnetic excitations for different momenta $\bm{q}$, segregating regions which promote $d$-wave pairing (near $\bm{q}=(\pi,\pi)$), are indifferent to pairing (along the AF zone boundary), and are hurtful to pairing (near zone center). The overall reduction of strength as well as hardening of magnetic spectral weight near $(\pi,\pi)$ leads to a decrease in the strength of the $d$-wave pair coupling consistent with the suppression of superconductivity in the overdoped regime.

The Hamiltonian for the Hubbard model appropriate for the hole doped cuprates
has the usual near neighbor hopping $t$, onsite $U$ and a negative next-near-neighbor
hopping $t'$.

\begin{equation}
  H = -t \sum_{\langle i j \rangle \sigma} c_{i\sigma}^\dagger c_{j\sigma}
      -t' \sum_{\langle\langle i j \rangle\rangle \sigma} c_{i\sigma}^\dagger c_{j\sigma}
      -\mu \sum_{i \sigma} n_{i\sigma} + U \sum_i n_{i\uparrow} n_{i\downarrow}
\label{eq:1}
\end{equation}
Here we will measure energies in units of $t$ and set $t'=-0.25$ and $U=6.5$.
The chemical potential $\mu$ in Eq.~(\ref{eq:1}) is used to fix the doping.
The DQMC calculations were carried out for an $8\times8$ lattice with 40 imaginary time slices
of width $\Delta\tau = 0.1$, for an inverse temperature of $\beta = 4.0$. For each doping level,
200 independently seeded Markov chains are run, each with $10^6$ full spacetime sweeps for
measurements.

\begin{figure}[htp]
  \includegraphics[width=7.5cm]{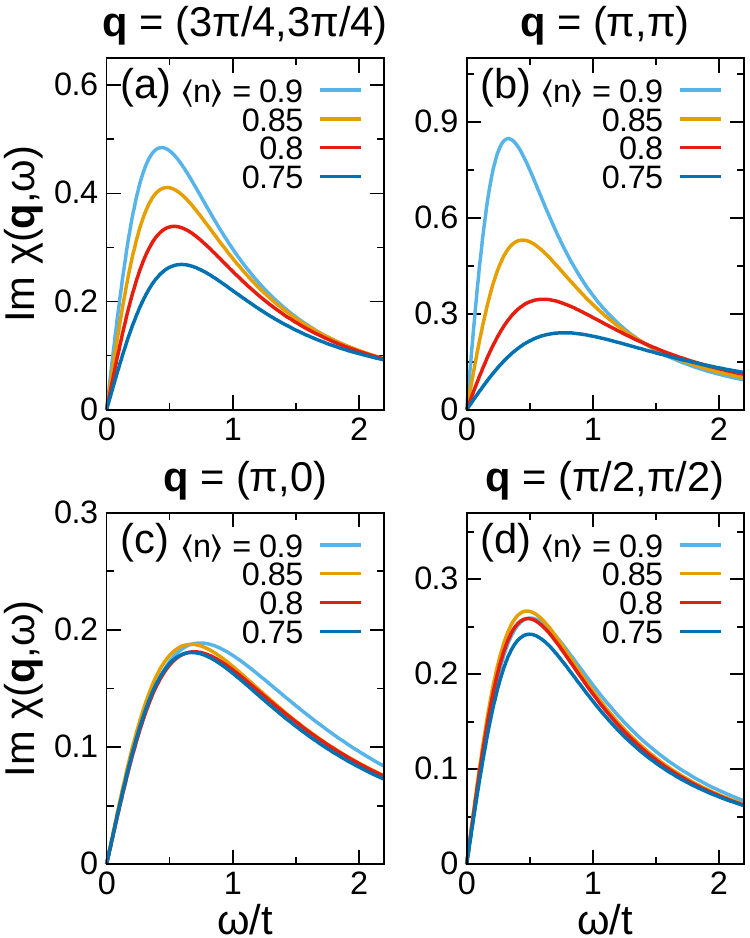}
\caption{The spin-fluctuation spectral weight ${\rm Im}\chi(\bm{q},\omega)$ versus
$\omega$ at a temperature $T=0.25t$ for several $\bm{q}$ values showing its evolution with doping. The
high energy magnetic excitations at the BZ boundary $\bm{q}=(\pi,0)$ and part way out along the
anti-nodal direction with $\bm{q}=(\pi/2,\pi/2)$ remain as the system is doped. However the spectral
weight associated with the magnetic excitations at larger anti-nodal momentum transfers
$\bm{q}=(3\pi/4,3\pi/4)$ and $(\pi,\pi)$ is reduced and shifted to higher frequencies.\label{fig:1}}
\end{figure}

The imaginary time spin susceptibility is calculated directly from DQMC as
\begin{equation}
\chi(\bm{q},\tau) = \sum_{\bm{r}} e^{-i \bm{q} \cdot \bm{r}} \frac{1}{N} \sum_{\bm{r'}} \left\langle S_z(\bm{r}+\bm{r'},\tau) S_z(\bm{r'},0)\right\rangle
\label{eq:2}
\end{equation}
where $S_z(\bm{r}) = \frac{1}{2}(n_{\bm{r}\uparrow} - n_{\bm{r}\downarrow})$ is the $z$ component of the spin at site $\bm{r}$. The real frequency
susceptibility is related to the imaginary time susceptibility by
\begin{equation}
\chi(\bm{q},\tau) = \int_0^\infty \frac{d\omega}{\pi} \frac{e^{-\tau\omega} + e^{-(\beta-\tau)\omega}}
{1 - e^{-\beta\omega}} {\rm Im} \chi(\bm{q},\omega).
\label{eq:chi}
\end{equation}
Since inverting Eq.~\ref{eq:chi} is numerically ill-posed, we use Maximum Entropy analytic continuation \cite{Jarrell1996}
to extract ${\rm Im}\chi(\bm{q},\omega)$ from the DQMC data. As described in Ref.~\cite{Jarrell1996,Gunnarsson2010}, a model function based on
the first moments of the data is used for the analytic continuation.

The spin fluctuation spectral weight ${\rm Im}\chi(\bm{q},\omega)$ for some selected
$\bm{q}$ values is plotted versus $\omega$ in Fig.~\ref{fig:1} for different dopings.
For the half-filled system, the $\bm{q}=(\pi,\pi)$ response continues to increase
and drop lower in frequency as $T$ decreases. However, for the doped system,
the spectral weight is well developed at this temperature and
the magnetic spin-fluctuation response evolves smoothly as the doping is increased.
For large momentum transfers near $(\pi,\pi)$, the hole doping both reduces
and shifts the spin-fluctuation spectral weight to higher frequencies. However, similar
to the RIXS data, for smaller anti-nodal momentum transfers $\bm{q}=(\pi/2,\pi/2)$ or for momentum
transfers along the nodal direction $\bm{q}=(\pi, 0)$, the peak in ${\rm Im}\chi(\bm{q},\omega)$ found in the DQMC calculations remains.

\begin{figure}[htp]
  \includegraphics[width=7.5cm]{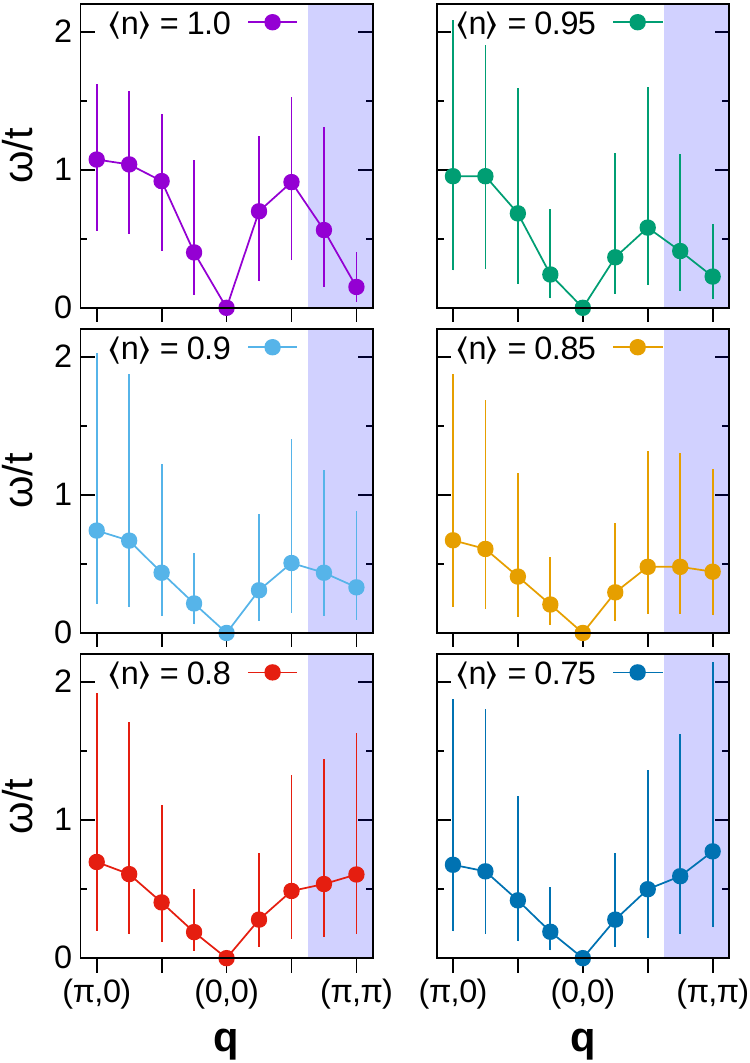}
\caption{The peak in the spin-fluctuation spectral weight ${\rm Im}\chi(\bm{q},\omega)$
versus $\bm{q}$ for different dopings. Here to the right of $(0,0)$, $\bm{q}$ moves along
the diagonal and to the left from $(0,0)$ to $(\pi,0)$. The shaded region at large momentum transfer
marks a region which is not measured by the RIXS experiments of Refs \cite{Dean2013,Meyers2017}.\label{fig:2}}
\end{figure}

To further illustrate the evolution of the calculated spin-fluctuation spectrum with doping, Fig.~\ref{fig:2}
shows a plot of the peak in ${\rm Im}\chi(\bm{q},\omega)$ for different dopings versus
$\bm{q}$ along the nodal and anti-nodal directions from zone center. The ends of the vertical bars mark
the energies where ${\rm Im}\chi(\bm{q},\omega)$ has dropped to half of its maximum value.
The unshaded region denotes the momentum transfer regime observed in the RIXS experiments.

From the results shown in Fig.~\ref{fig:1} and \ref{fig:2}, one can see that while
doping leads to changes in the overall magnetic excitation spectrum, the AF excitations accessible to RIXS remain relatively unchanged with doping. There is a clear similarity between the experimental RIXS
data for LSCO and the DQMC results. However, the region outside of the reach of transition metal $L-$edge RIXS near $(\pi,\pi)$, due to the overall scale of photon momenta, changes considerably and, as we will discuss, has an impact on the strength of $d$-wave pairing in the Hubbard model.

\begin{figure}[tp]
  \includegraphics[width=7.5cm]{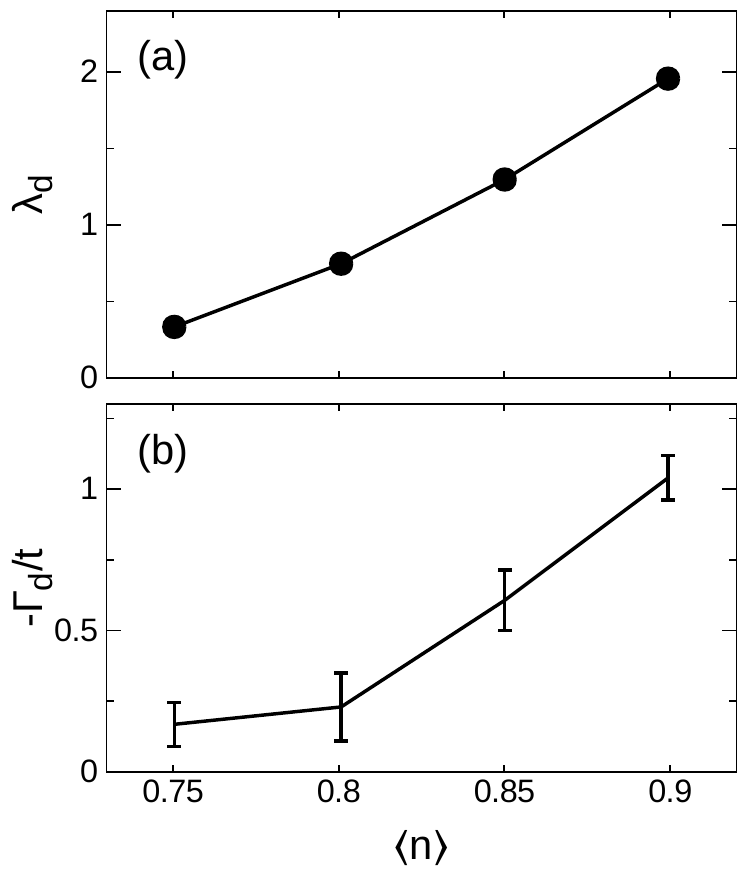}
\caption{(a) The strength $\lambda_d$ of the d-wave pairing interaction given by Eq.~(\ref{eq:lambda_d}) , versus doping at $T=0.25t$ (b) The interaction vertex $\Gamma_d$, Eq.~(\ref{eq:gammad}), versus doping at $T=0.25t$.\label{fig:3}}
\end{figure}

A measure of the strength of the spin-fluctuation $d$-wave pairing interaction in weak coupling \cite{Maier2007} is given by
\begin{equation}
  \lambda_d=-\frac{3}{2}U^2\left\langle\phi_d(\bm{k})\int^\infty_0\frac{d\omega}{\pi}
	\frac{{\rm Im}\chi(\bm{k}-\bm{k'},\omega)}{\omega}\phi_d(\bm{k'})\right\rangle_{\rm FS}\Big/\left\langle\phi^2_d(\bm{k})\right\rangle_{\rm FS}
\label{eq:lambda_d}
\end{equation}
Here $\phi_d(\bm{k})=(\cos k_x-\cos k_y)$ and the $\bm{k}$ averages are taken over a region
of band energies $\pm0.5t$ around the Fermi surface. A plot of $\lambda_d$ versus
doping is shown in Fig.~\ref{fig:3}a. Here one sees that this coupling strength decreases with doping.
This same behavior is observed in a direct calculation of the correlated and uncorrelated $d$-wave pair-field susceptibilities \cite{White1989,Moreo1991} and the corresponding interaction vertex, defined respectively as
\begin{equation}
  P_d = \int_0^\beta d\tau \frac{1}{N^2} \sum_{\bm{k},\bm{k'}} \phi_d(\bm{k})
  \left\langle 
    c_{-\bm{k}\downarrow}(\tau)
    c_{\bm{k}\uparrow}(\tau)
    c_{\bm{k'}\uparrow}^{\dagger}(0)
    c_{-\bm{k'}\downarrow}^{\dagger}(0)
  \right\rangle \phi_d(\bm{k'})
\label{eq:pd}
\end{equation}
\begin{equation}
  \overline{P}_d = \int_0^\beta d\tau \frac{1}{N^2} \sum_{\bm{k}} \phi_d^2(\bm{k})
  \left\langle 
    c_{-\bm{k}\downarrow}(\tau)
    c_{-\bm{k}\downarrow}^{\dagger}(0)
  \right\rangle
  \left\langle
    c_{\bm{k}\uparrow}(\tau)
    c_{\bm{k}\uparrow}^{\dagger}(0)
  \right\rangle
\label{eq:pdbar}
\end{equation}
\begin{equation}
  \Gamma_d = \frac{1}{P_d} - \frac{1}{\overline{P}_d}
\label{eq:gammad}
\end{equation}
The interaction vertex $\Gamma_d$ provides another gauge of the $d$-wave pairing strength, with negative values indicating an attractive interaction. As plotted in Fig.~\ref{fig:3}b, this measure confirms the decrease of the pairing interaction upon doping similar to the behavior of $\lambda_d$ seen in Fig.~\ref{fig:3}a.
The decrease of both $\lambda_d$ and $-\Gamma_d$ reflects the reduction and hardening of the
spin-fluctuation spectral weight in the large momentum $\bm{q} \sim (\pi,\pi)$ transfer region marked by
the shaded regions of Fig.~\ref{fig:2}.

\begin{figure}
  \includegraphics[width=7.5cm]{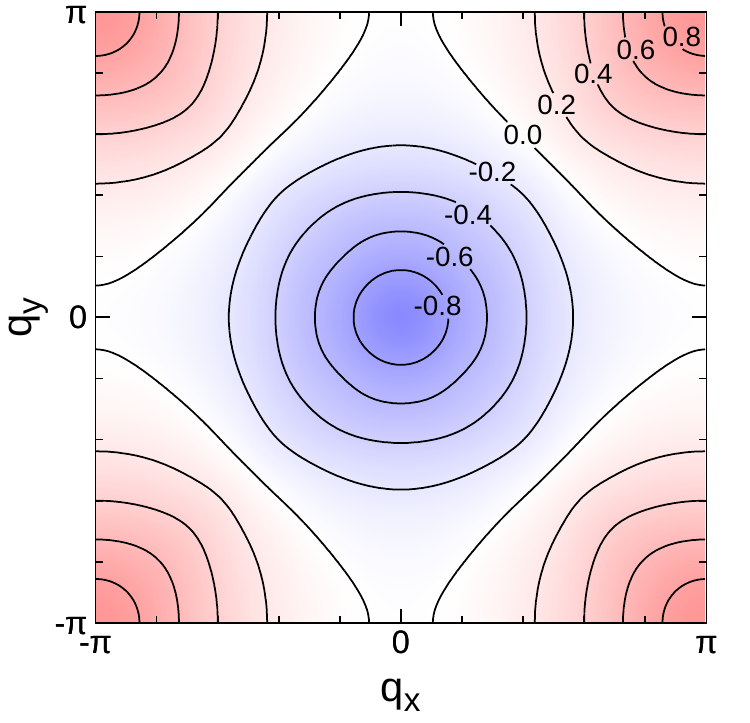}
\caption{Plot of $F(\bm{q})$, Eq.~(\ref{eq:F}), normalized to its absolute value at $\bm{q} = (0,0)$, versus $(q_x,q_y)$ over the first Brillouin zone using
the same $\phi_d(\bm{k})$ gap functions and cut-off around the Fermi surface as in Fig.~\ref{fig:3}.
Momentum transfers near $(\pi,\pi)$ (red shaded region) lead to a positive contribution from the spin-fluctuations to the coupling strength $\lambda_d$. Spin-fluctuations with momentum transfers near $(0,0)$ (blue shaded region) give a negative contribution.\label{fig:4}
}
\end{figure}

The high energy magnetic excitations seen by RIXS at the edge of the BZ in the
anti-nodal direction as well those seen along the nodal direction with $\bm{q} = (\pi/2,\pi/2)$ lack the momentum transfer to scatter pairs between the anti-nodal regions. This is illustrated in Fig.~\ref{fig:4} which shows a plot of the convolved $d$-wave form factor
\begin{equation}
  F(\bm{q})=-\left\langle\phi_d(\bm{k})\phi_d(\bm{k}+\bm{q})\right\rangle_{\rm FS}
\label{eq:F}
\end{equation}
for $\langle n \rangle = 0.9$. The pairing strength $\lambda_d$ given by Eq.~(\ref{eq:lambda_d}) is proportional to
a weighted average of ${\rm Re}\chi(\bm{q},\omega=0)$ with respect to $F(\bm{q})$. As shown in Fig.~\ref{fig:4},
the spectral weight of the spin-fluctuations at large momenta transfer give rise to the $d$-wave
pairing while the small momentum transfers suppress the pairing. The intermediate region, where the RIXS
experiments find magnetic excitations, play a marginal role as earlier suggested in Ref \cite{Meyers2017}.
Regions near $(\pi,\pi)$ which are accessible via polarized inelastic neutron scattering provide the dominant contribution to the strength of the d-wave pairing interaction. A closer inspection of the momentum regions accessible near the zone center by RIXS would also be useful in understanding the decrease of pairing strength. The DQMC results reported here and elsewhere are consistent with the weakening of spectral intensity and hardening of the spin excitations observed near the magnetic zone center $(\pi,\pi)$ in those measurements \cite{Wakimoto2004,Wakimoto2005,Wakimoto2007,Lipscombe2007,Wakimoto2015}. Thus we conclude that the evolution of the spin spectrum of excitations with doping in the Hubbard model is consistent with the existing data in the cuprates and can account for the reduction of $d$-wave pairing strength with doping.

\section*{Acknowledgments}

We acknowledge helpful discussions with Christian Mendl, Yvonne Kung, and Chunjing Jia. Computational work was performed on the Sherlock cluster at Stanford University. This work was supported by the U.S.~Department of Energy, Office of Basic Energy Sciences, Division of Materials Sciences and Engineering, under Contract No.~DE-AC02-76SF00515. A portion of this research was conducted at the Center for Nanophase Materials Sciences, which is a DOE Office of Science User Facility.

\bibliographystyle{apsrev4-1}

\begin{thebibliography}{16}%
\makeatletter
\providecommand \@ifxundefined [1]{%
 \@ifx{#1\undefined}
}%
\providecommand \@ifnum [1]{%
 \ifnum #1\expandafter \@firstoftwo
 \else \expandafter \@secondoftwo
 \fi
}%
\providecommand \@ifx [1]{%
 \ifx #1\expandafter \@firstoftwo
 \else \expandafter \@secondoftwo
 \fi
}%
\providecommand \natexlab [1]{#1}%
\providecommand \enquote  [1]{``#1''}%
\providecommand \bibnamefont  [1]{#1}%
\providecommand \bibfnamefont [1]{#1}%
\providecommand \citenamefont [1]{#1}%
\providecommand \href@noop [0]{\@secondoftwo}%
\providecommand \href [0]{\begingroup \@sanitize@url \@href}%
\providecommand \@href[1]{\@@startlink{#1}\@@href}%
\providecommand \@@href[1]{\endgroup#1\@@endlink}%
\providecommand \@sanitize@url [0]{\catcode `\\12\catcode `\$12\catcode
  `\&12\catcode `\#12\catcode `\^12\catcode `\_12\catcode `\%12\relax}%
\providecommand \@@startlink[1]{}%
\providecommand \@@endlink[0]{}%
\providecommand \url  [0]{\begingroup\@sanitize@url \@url }%
\providecommand \@url [1]{\endgroup\@href {#1}{\urlprefix }}%
\providecommand \urlprefix  [0]{URL }%
\providecommand \Eprint [0]{\href }%
\providecommand \doibase [0]{http://dx.doi.org/}%
\providecommand \selectlanguage [0]{\@gobble}%
\providecommand \bibinfo  [0]{\@secondoftwo}%
\providecommand \bibfield  [0]{\@secondoftwo}%
\providecommand \translation [1]{[#1]}%
\providecommand \BibitemOpen [0]{}%
\providecommand \bibitemStop [0]{}%
\providecommand \bibitemNoStop [0]{.\EOS\space}%
\providecommand \EOS [0]{\spacefactor3000\relax}%
\providecommand \BibitemShut  [1]{\csname bibitem#1\endcsname}%
\let\auto@bib@innerbib\@empty
\bibitem [{\citenamefont {Dean}\ \emph {et~al.}(2013)\citenamefont {Dean},
  \citenamefont {Dellea}, \citenamefont {Springell}, \citenamefont
  {Yakhou-Harris}, \citenamefont {Kummer}, \citenamefont {Brookes},
  \citenamefont {Liu}, \citenamefont {Sun}, \citenamefont {Strle},
  \citenamefont {Schmitt}, \citenamefont {Braicovich}, \citenamefont
  {Ghiringhelli}, \citenamefont {Bo\ifmmode \check{z}\else
  \v{z}\fi{}ovi\ifmmode~\acute{c}\else \'{c}\fi{}},\ and\ \citenamefont
  {Hill}}]{Dean2013}%
  \BibitemOpen
  \bibfield  {author} {\bibinfo {author} {\bibfnamefont {M.~P.~M.}\
  \bibnamefont {Dean}}, \bibinfo {author} {\bibfnamefont {G.}~\bibnamefont
  {Dellea}}, \bibinfo {author} {\bibfnamefont {R.~S.}\ \bibnamefont
  {Springell}}, \bibinfo {author} {\bibfnamefont {F.}~\bibnamefont
  {Yakhou-Harris}}, \bibinfo {author} {\bibfnamefont {K.}~\bibnamefont
  {Kummer}}, \bibinfo {author} {\bibfnamefont {N.~B.}\ \bibnamefont {Brookes}},
  \bibinfo {author} {\bibfnamefont {X.}~\bibnamefont {Liu}}, \bibinfo {author}
  {\bibfnamefont {Y.-J.}\ \bibnamefont {Sun}}, \bibinfo {author} {\bibfnamefont
  {J.}~\bibnamefont {Strle}}, \bibinfo {author} {\bibfnamefont
  {T.}~\bibnamefont {Schmitt}}, \bibinfo {author} {\bibfnamefont
  {L.}~\bibnamefont {Braicovich}}, \bibinfo {author} {\bibfnamefont
  {G.}~\bibnamefont {Ghiringhelli}}, \bibinfo {author} {\bibfnamefont
  {I.}~\bibnamefont {Bo\ifmmode \check{z}\else
  \v{z}\fi{}ovi\ifmmode~\acute{c}\else \'{c}\fi{}}}, \ and\ \bibinfo {author}
  {\bibfnamefont {J.~P.}\ \bibnamefont {Hill}},\ }\href {\doibase
  10.1038/nmat3723} {\bibfield  {journal} {\bibinfo  {journal} {Nat. Mater.}\
  }\textbf {\bibinfo {volume} {12}},\ \bibinfo {pages} {1019} (\bibinfo {year}
  {2013})}\BibitemShut {NoStop}%
\bibitem [{\citenamefont {Wakimoto}\ \emph {et~al.}(2015)\citenamefont
  {Wakimoto}, \citenamefont {Ishii}, \citenamefont {Kimura}, \citenamefont
  {Fujita}, \citenamefont {Dellea}, \citenamefont {Kummer}, \citenamefont
  {Braicovich}, \citenamefont {Ghiringhelli}, \citenamefont {Debeer-Schmitt},\
  and\ \citenamefont {Granroth}}]{Wakimoto2015}%
  \BibitemOpen
  \bibfield  {author} {\bibinfo {author} {\bibfnamefont {S.}~\bibnamefont
  {Wakimoto}}, \bibinfo {author} {\bibfnamefont {K.}~\bibnamefont {Ishii}},
  \bibinfo {author} {\bibfnamefont {H.}~\bibnamefont {Kimura}}, \bibinfo
  {author} {\bibfnamefont {M.}~\bibnamefont {Fujita}}, \bibinfo {author}
  {\bibfnamefont {G.}~\bibnamefont {Dellea}}, \bibinfo {author} {\bibfnamefont
  {K.}~\bibnamefont {Kummer}}, \bibinfo {author} {\bibfnamefont
  {L.}~\bibnamefont {Braicovich}}, \bibinfo {author} {\bibfnamefont
  {G.}~\bibnamefont {Ghiringhelli}}, \bibinfo {author} {\bibfnamefont {L.~M.}\
  \bibnamefont {Debeer-Schmitt}}, \ and\ \bibinfo {author} {\bibfnamefont
  {G.~E.}\ \bibnamefont {Granroth}},\ }\href {\doibase
  10.1103/PhysRevB.91.184513} {\bibfield  {journal} {\bibinfo  {journal} {Phys.
  Rev. B}\ }\textbf {\bibinfo {volume} {91}},\ \bibinfo {pages} {184513}
  (\bibinfo {year} {2015})}\BibitemShut {NoStop}%
\bibitem [{\citenamefont {Meyers}\ \emph {et~al.}(2017)\citenamefont {Meyers},
  \citenamefont {Miao}, \citenamefont {Walters}, \citenamefont {Bisogni},
  \citenamefont {Springell}, \citenamefont {d'Astuto}, \citenamefont {Dantz},
  \citenamefont {Pelliciari}, \citenamefont {Huang}, \citenamefont {Okamoto},
  \citenamefont {Huang}, \citenamefont {Hill}, \citenamefont {He},
  \citenamefont {Bo\ifmmode \check{z}\else \v{z}\fi{}ovi\ifmmode~\acute{c}\else
  \'{c}\fi{}}, \citenamefont {Schmitt},\ and\ \citenamefont
  {Dean}}]{Meyers2017}%
  \BibitemOpen
  \bibfield  {author} {\bibinfo {author} {\bibfnamefont {D.}~\bibnamefont
  {Meyers}}, \bibinfo {author} {\bibfnamefont {H.}~\bibnamefont {Miao}},
  \bibinfo {author} {\bibfnamefont {A.~C.}\ \bibnamefont {Walters}}, \bibinfo
  {author} {\bibfnamefont {V.}~\bibnamefont {Bisogni}}, \bibinfo {author}
  {\bibfnamefont {R.~S.}\ \bibnamefont {Springell}}, \bibinfo {author}
  {\bibfnamefont {M.}~\bibnamefont {d'Astuto}}, \bibinfo {author}
  {\bibfnamefont {M.}~\bibnamefont {Dantz}}, \bibinfo {author} {\bibfnamefont
  {J.}~\bibnamefont {Pelliciari}}, \bibinfo {author} {\bibfnamefont {H.~Y.}\
  \bibnamefont {Huang}}, \bibinfo {author} {\bibfnamefont {J.}~\bibnamefont
  {Okamoto}}, \bibinfo {author} {\bibfnamefont {D.~J.}\ \bibnamefont {Huang}},
  \bibinfo {author} {\bibfnamefont {J.~P.}\ \bibnamefont {Hill}}, \bibinfo
  {author} {\bibfnamefont {X.}~\bibnamefont {He}}, \bibinfo {author}
  {\bibfnamefont {I.}~\bibnamefont {Bo\ifmmode \check{z}\else
  \v{z}\fi{}ovi\ifmmode~\acute{c}\else \'{c}\fi{}}}, \bibinfo {author}
  {\bibfnamefont {T.}~\bibnamefont {Schmitt}}, \ and\ \bibinfo {author}
  {\bibfnamefont {M.~P.~M.}\ \bibnamefont {Dean}},\ }\href {\doibase
  10.1103/PhysRevB.95.075139} {\bibfield  {journal} {\bibinfo  {journal} {Phys.
  Rev. B}\ }\textbf {\bibinfo {volume} {95}},\ \bibinfo {pages} {075139}
  (\bibinfo {year} {2017})}\BibitemShut {NoStop}%
\bibitem [{\citenamefont {Scalapino}(2012)}]{Scalapino2012}%
  \BibitemOpen
  \bibfield  {author} {\bibinfo {author} {\bibfnamefont {D.~J.}\ \bibnamefont
  {Scalapino}},\ }\href {\doibase 10.1103/RevModPhys.84.1383} {\bibfield
  {journal} {\bibinfo  {journal} {Rev. Mod. Phys.}\ }\textbf {\bibinfo {volume}
  {84}},\ \bibinfo {pages} {1383} (\bibinfo {year} {2012})}\BibitemShut
  {NoStop}%
\bibitem [{\citenamefont {White}\ \emph {et~al.}(1989)\citenamefont {White},
  \citenamefont {Scalapino}, \citenamefont {Sugar}, \citenamefont {Loh},
  \citenamefont {Gubernatis},\ and\ \citenamefont {Scalettar}}]{White1989}%
  \BibitemOpen
  \bibfield  {author} {\bibinfo {author} {\bibfnamefont {S.~R.}\ \bibnamefont
  {White}}, \bibinfo {author} {\bibfnamefont {D.~J.}\ \bibnamefont
  {Scalapino}}, \bibinfo {author} {\bibfnamefont {R.~L.}\ \bibnamefont
  {Sugar}}, \bibinfo {author} {\bibfnamefont {E.~Y.}\ \bibnamefont {Loh}},
  \bibinfo {author} {\bibfnamefont {J.~E.}\ \bibnamefont {Gubernatis}}, \ and\
  \bibinfo {author} {\bibfnamefont {R.~T.}\ \bibnamefont {Scalettar}},\ }\href
  {\doibase 10.1103/PhysRevB.40.506} {\bibfield  {journal} {\bibinfo  {journal}
  {Phys. Rev. B}\ }\textbf {\bibinfo {volume} {40}},\ \bibinfo {pages} {506}
  (\bibinfo {year} {1989})}\BibitemShut {NoStop}%
\bibitem [{\citenamefont {Jia}\ \emph {et~al.}(2014)\citenamefont {Jia},
  \citenamefont {Nowadnick}, \citenamefont {Wohlfeld}, \citenamefont {Kung},
  \citenamefont {Chen}, \citenamefont {Johnston}, \citenamefont {Tohyama},
  \citenamefont {Moritz},\ and\ \citenamefont {Devereaux}}]{Jia2014}%
  \BibitemOpen
  \bibfield  {author} {\bibinfo {author} {\bibfnamefont {C.~J.}\ \bibnamefont
  {Jia}}, \bibinfo {author} {\bibfnamefont {E.~A.}\ \bibnamefont {Nowadnick}},
  \bibinfo {author} {\bibfnamefont {K.}~\bibnamefont {Wohlfeld}}, \bibinfo
  {author} {\bibfnamefont {Y.~F.}\ \bibnamefont {Kung}}, \bibinfo {author}
  {\bibfnamefont {C.-C.}\ \bibnamefont {Chen}}, \bibinfo {author}
  {\bibfnamefont {S.}~\bibnamefont {Johnston}}, \bibinfo {author}
  {\bibfnamefont {T.}~\bibnamefont {Tohyama}}, \bibinfo {author} {\bibfnamefont
  {B.}~\bibnamefont {Moritz}}, \ and\ \bibinfo {author} {\bibfnamefont {T.~P.}\
  \bibnamefont {Devereaux}},\ }\href {\doibase 10.1038/ncomms4314} {\bibfield
  {journal} {\bibinfo  {journal} {Nature Commun.}\ }\textbf {\bibinfo {volume}
  {5}},\ \bibinfo {pages} {3314} (\bibinfo {year} {2014})}\BibitemShut
  {NoStop}%
\bibitem [{\citenamefont {Kung}\ \emph {et~al.}(2015)\citenamefont {Kung},
  \citenamefont {Nowadnick}, \citenamefont {Jia}, \citenamefont {Johnston},
  \citenamefont {Moritz}, \citenamefont {Scalettar},\ and\ \citenamefont
  {Devereaux}}]{Kung2015}%
  \BibitemOpen
  \bibfield  {author} {\bibinfo {author} {\bibfnamefont {Y.~F.}\ \bibnamefont
  {Kung}}, \bibinfo {author} {\bibfnamefont {E.~A.}\ \bibnamefont {Nowadnick}},
  \bibinfo {author} {\bibfnamefont {C.~J.}\ \bibnamefont {Jia}}, \bibinfo
  {author} {\bibfnamefont {S.}~\bibnamefont {Johnston}}, \bibinfo {author}
  {\bibfnamefont {B.}~\bibnamefont {Moritz}}, \bibinfo {author} {\bibfnamefont
  {R.~T.}\ \bibnamefont {Scalettar}}, \ and\ \bibinfo {author} {\bibfnamefont
  {T.~P.}\ \bibnamefont {Devereaux}},\ }\href {\doibase
  10.1103/PhysRevB.92.195108} {\bibfield  {journal} {\bibinfo  {journal} {Phys.
  Rev. B}\ }\textbf {\bibinfo {volume} {92}},\ \bibinfo {pages} {195108}
  (\bibinfo {year} {2015})}\BibitemShut {NoStop}%
\bibitem [{\citenamefont {Ament}\ \emph {et~al.}(2011)\citenamefont {Ament},
  \citenamefont {van Veenendaal}, \citenamefont {Devereaux}, \citenamefont
  {Hill},\ and\ \citenamefont {van~den Brink}}]{Ament2011}%
  \BibitemOpen
  \bibfield  {author} {\bibinfo {author} {\bibfnamefont {L.~J.~P.}\
  \bibnamefont {Ament}}, \bibinfo {author} {\bibfnamefont {M.}~\bibnamefont
  {van Veenendaal}}, \bibinfo {author} {\bibfnamefont {T.~P.}\ \bibnamefont
  {Devereaux}}, \bibinfo {author} {\bibfnamefont {J.~P.}\ \bibnamefont {Hill}},
  \ and\ \bibinfo {author} {\bibfnamefont {J.}~\bibnamefont {van~den Brink}},\
  }\href {\doibase 10.1103/RevModPhys.83.705} {\bibfield  {journal} {\bibinfo
  {journal} {Rev. Mod. Phys.}\ }\textbf {\bibinfo {volume} {83}},\ \bibinfo
  {pages} {705} (\bibinfo {year} {2011})}\BibitemShut {NoStop}%
\bibitem [{\citenamefont {Jarrell}\ and\ \citenamefont
  {Gubernatis}(1996)}]{Jarrell1996}%
  \BibitemOpen
  \bibfield  {author} {\bibinfo {author} {\bibfnamefont {M.}~\bibnamefont
  {Jarrell}}\ and\ \bibinfo {author} {\bibfnamefont {J.~E.}\ \bibnamefont
  {Gubernatis}},\ }\href {\doibase 10.1016/0370-1573(95)00074-7} {\bibfield
  {journal} {\bibinfo  {journal} {Phys. Rep.}\ }\textbf {\bibinfo {volume}
  {269}},\ \bibinfo {pages} {133} (\bibinfo {year} {1996})}\BibitemShut
  {NoStop}%
\bibitem [{\citenamefont {Gunnarsson}\ \emph {et~al.}(2010)\citenamefont
  {Gunnarsson}, \citenamefont {Haverkort},\ and\ \citenamefont
  {Sangiovanni}}]{Gunnarsson2010}%
  \BibitemOpen
  \bibfield  {author} {\bibinfo {author} {\bibfnamefont {O.}~\bibnamefont
  {Gunnarsson}}, \bibinfo {author} {\bibfnamefont {M.~W.}\ \bibnamefont
  {Haverkort}}, \ and\ \bibinfo {author} {\bibfnamefont {G.}~\bibnamefont
  {Sangiovanni}},\ }\href {\doibase 10.1103/PhysRevB.81.155107} {\bibfield
  {journal} {\bibinfo  {journal} {Phys. Rev. B}\ }\textbf {\bibinfo {volume}
  {81}},\ \bibinfo {pages} {155107} (\bibinfo {year} {2010})}\BibitemShut
  {NoStop}%
\bibitem [{\citenamefont {Maier}\ \emph {et~al.}(2007)\citenamefont {Maier},
  \citenamefont {Jarrell},\ and\ \citenamefont {Scalapino}}]{Maier2007}%
  \BibitemOpen
  \bibfield  {author} {\bibinfo {author} {\bibfnamefont {T.~A.}\ \bibnamefont
  {Maier}}, \bibinfo {author} {\bibfnamefont {M.}~\bibnamefont {Jarrell}}, \
  and\ \bibinfo {author} {\bibfnamefont {D.~J.}\ \bibnamefont {Scalapino}},\
  }\href {\doibase 10.1103/PhysRevB.75.134519} {\bibfield  {journal} {\bibinfo
  {journal} {Phys. Rev. B}\ }\textbf {\bibinfo {volume} {75}},\ \bibinfo
  {pages} {134519} (\bibinfo {year} {2007})}\BibitemShut {NoStop}%
\bibitem [{\citenamefont {Moreo}\ and\ \citenamefont
  {Scalapino}(1991)}]{Moreo1991}%
  \BibitemOpen
  \bibfield  {author} {\bibinfo {author} {\bibfnamefont {A.}~\bibnamefont
  {Moreo}}\ and\ \bibinfo {author} {\bibfnamefont {D.~J.}\ \bibnamefont
  {Scalapino}},\ }\href {\doibase 10.1103/PhysRevB.43.8211} {\bibfield
  {journal} {\bibinfo  {journal} {Phys. Rev. B}\ }\textbf {\bibinfo {volume}
  {43}},\ \bibinfo {pages} {8211} (\bibinfo {year} {1991})}\BibitemShut
  {NoStop}%
\bibitem [{\citenamefont {Wakimoto}\ \emph {et~al.}(2004)\citenamefont
  {Wakimoto}, \citenamefont {Zhang}, \citenamefont {Yamada}, \citenamefont
  {Swainson}, \citenamefont {Kim},\ and\ \citenamefont
  {Birgeneau}}]{Wakimoto2004}%
  \BibitemOpen
  \bibfield  {author} {\bibinfo {author} {\bibfnamefont {S.}~\bibnamefont
  {Wakimoto}}, \bibinfo {author} {\bibfnamefont {H.}~\bibnamefont {Zhang}},
  \bibinfo {author} {\bibfnamefont {K.}~\bibnamefont {Yamada}}, \bibinfo
  {author} {\bibfnamefont {I.}~\bibnamefont {Swainson}}, \bibinfo {author}
  {\bibfnamefont {H.}~\bibnamefont {Kim}}, \ and\ \bibinfo {author}
  {\bibfnamefont {R.~J.}\ \bibnamefont {Birgeneau}},\ }\href {\doibase
  10.1103/PhysRevLett.92.217004} {\bibfield  {journal} {\bibinfo  {journal}
  {Phys. Rev. Lett.}\ }\textbf {\bibinfo {volume} {92}},\ \bibinfo {pages}
  {217004} (\bibinfo {year} {2004})}\BibitemShut {NoStop}%
\bibitem [{\citenamefont {Wakimoto}\ \emph {et~al.}(2005)\citenamefont
  {Wakimoto}, \citenamefont {Birgeneau}, \citenamefont {Kagedan}, \citenamefont
  {Kim}, \citenamefont {Swainson}, \citenamefont {Yamada},\ and\ \citenamefont
  {Zhang}}]{Wakimoto2005}%
  \BibitemOpen
  \bibfield  {author} {\bibinfo {author} {\bibfnamefont {S.}~\bibnamefont
  {Wakimoto}}, \bibinfo {author} {\bibfnamefont {R.~J.}\ \bibnamefont
  {Birgeneau}}, \bibinfo {author} {\bibfnamefont {A.}~\bibnamefont {Kagedan}},
  \bibinfo {author} {\bibfnamefont {H.}~\bibnamefont {Kim}}, \bibinfo {author}
  {\bibfnamefont {I.}~\bibnamefont {Swainson}}, \bibinfo {author}
  {\bibfnamefont {K.}~\bibnamefont {Yamada}}, \ and\ \bibinfo {author}
  {\bibfnamefont {H.}~\bibnamefont {Zhang}},\ }\href {\doibase
  10.1103/PhysRevB.72.064521} {\bibfield  {journal} {\bibinfo  {journal} {Phys.
  Rev. B}\ }\textbf {\bibinfo {volume} {72}},\ \bibinfo {pages} {064521}
  (\bibinfo {year} {2005})}\BibitemShut {NoStop}%
\bibitem [{\citenamefont {Wakimoto}\ \emph {et~al.}(2007)\citenamefont
  {Wakimoto}, \citenamefont {Yamada}, \citenamefont {Tranquada}, \citenamefont
  {Frost}, \citenamefont {Birgeneau},\ and\ \citenamefont
  {Zhang}}]{Wakimoto2007}%
  \BibitemOpen
  \bibfield  {author} {\bibinfo {author} {\bibfnamefont {S.}~\bibnamefont
  {Wakimoto}}, \bibinfo {author} {\bibfnamefont {K.}~\bibnamefont {Yamada}},
  \bibinfo {author} {\bibfnamefont {J.~M.}\ \bibnamefont {Tranquada}}, \bibinfo
  {author} {\bibfnamefont {C.~D.}\ \bibnamefont {Frost}}, \bibinfo {author}
  {\bibfnamefont {R.~J.}\ \bibnamefont {Birgeneau}}, \ and\ \bibinfo {author}
  {\bibfnamefont {H.}~\bibnamefont {Zhang}},\ }\href {\doibase
  10.1103/PhysRevLett.98.247003} {\bibfield  {journal} {\bibinfo  {journal}
  {Phys. Rev. Lett.}\ }\textbf {\bibinfo {volume} {98}},\ \bibinfo {pages}
  {247003} (\bibinfo {year} {2007})}\BibitemShut {NoStop}%
\bibitem [{\citenamefont {Lipscombe}\ \emph {et~al.}(2007)\citenamefont
  {Lipscombe}, \citenamefont {Hayden}, \citenamefont {Vignolle}, \citenamefont
  {McMorrow},\ and\ \citenamefont {Perring}}]{Lipscombe2007}%
  \BibitemOpen
  \bibfield  {author} {\bibinfo {author} {\bibfnamefont {O.~J.}\ \bibnamefont
  {Lipscombe}}, \bibinfo {author} {\bibfnamefont {S.~M.}\ \bibnamefont
  {Hayden}}, \bibinfo {author} {\bibfnamefont {B.}~\bibnamefont {Vignolle}},
  \bibinfo {author} {\bibfnamefont {D.~F.}\ \bibnamefont {McMorrow}}, \ and\
  \bibinfo {author} {\bibfnamefont {T.~G.}\ \bibnamefont {Perring}},\ }\href
  {\doibase 10.1103/PhysRevLett.99.067002} {\bibfield  {journal} {\bibinfo
  {journal} {Phys. Rev. Lett.}\ }\textbf {\bibinfo {volume} {99}},\ \bibinfo
  {pages} {067002} (\bibinfo {year} {2007})}\BibitemShut {NoStop}%
\end{thebibliography}
%

\end{document}